\documentclass[aip, amsmath,amssymb, reprint]{revtex4-1}

\usepackage{graphicx}
\usepackage{tabularx}
\usepackage{color}
\usepackage{amsmath}
\usepackage{tikz}
\usetikzlibrary{tikzmark,fit}
\usepackage{comment}
\newcommand{\bea}{\begin{eqnarray}}
\newcommand{\eea}{\end{eqnarray}} 
\usepackage{float}
\usepackage{braket}
\usepackage{url}

\newcommand{\be}{\begin{equation}} 
\newcommand{\ee}{\end{equation}}

\newcommand{\tn}{\tau}
\usepackage{dcolumn}
\usepackage{hyperref}
\usepackage{bm}
\usepackage{epsf}
\usepackage{subfigure}
\usepackage{epstopdf}%
\usepackage{amsfonts}

\DeclareMathOperator{\trace}{Tr}

\begin{document}

\title{Long-Range Charge Transport in Homogeneous and Alternating-Rigidity Chains}


\author{Francisco Lai Liang}
\email{francisco.lai@mail.utoronto.ca}
\affiliation{Department of Chemistry and Centre for Quantum Information and Quantum Control,
University of Toronto, 80 Saint George St., Toronto, Ontario, Canada M5S 3H6}
\author{Dvira Segal}
\email{dvira.segal@utoronto.ca}
\affiliation{Department of Chemistry and Centre for Quantum Information and Quantum Control,
University of Toronto, 80 Saint George St., Toronto, Ontario, Canada M5S 3H6}
\affiliation{Department of Physics, University of Toronto, Toronto, Ontario, Canada M5S 1A7}

\date{\today}
\begin{abstract}
We study the interplay of intrinsic-electronic and environmental factors on
long-range charge transport across molecular chains with up to $N\sim 80$ monomers.
We describe the molecular electronic structure of the chain with a tight-binding Hamiltonian.
Thermal effects in the form of electron decoherence and inelastic scatterings
are incorporated with the Landauer-B\"uttiker probe method.
In short chains of up to 10 units we observe the crossover between coherent
(tunneling, ballistic) motion and thermally-assisted conduction, 
with thermal effects enhancing the current beyond the quantum coherent limit.
We further show that unconventional (nonmonotonic with size) transport behavior emerges 
when monomer-to-monomer electronic coupling is made large. 
In long chains, we identify a different behavior, with 
thermal effects suppressing the conductance below the coherent-ballistic limit.
With the goal to identify a minimal model for molecular chains displaying unconventional and
effective long-range transport, we simulate a modular polymer
with alternating regions of high and low rigidity.
Simulations show that, surprisingly, while charge correlations are significantly affected by 
structuring environmental conditions, reflecting charge delocalization,
the electrical resistance displays an averaging effect, 
and it is not sensitive to this patterning.
We conclude by arguing that efficient long-range charge transport 
requires engineering both internal electronic parameters and environmental conditions. 
\end{abstract}

\maketitle 

\section{Introduction}
\label{Sec-intro}

Electron transport is central to biology with many life-sustaining processes relying on it 
including photosynthesis, protein function, DNA mutagenesis and carcinogenesis \cite{Barton}. 
Furthermore, charge transport is obviously at the heart of technologies such as electronics, 
thermoelectric energy generation, optoelectronics, spintronics, and plasmonics.
Atomic-scale and single-molecule junctions represent the ultimate limit of 
miniaturization of electronic devices. They further allow unique and fundamental studies, 
such as the detection of chemical reactions at the resolution of a single molecule in time \cite{Guo2021}.
 
Experimental studies of charge transport across ultrashort metal-molecule-metal junctions,
shorter than 1 nm  (made e.g., of a single benzene ring) 
are well explained with the fully quantum-coherent Landauer approach \cite{Cuevas2010}, see for example
Ref. \cite{Latha}.
However, longer molecules such as oligomers with up to 10 units and extending up to 10 nm, 
demonstrate the onset of thermally-assisted transport, with 
thermal effects in the electrodes and the molecule facilitating 
conduction \cite{Frisbie1,Frisbie2,Ratner, Mccree, Tao,Lacroix,Carmen}. 
Indeed, based on timescale considerations,
in long molecules, the impact of nuclear degrees of freedom on electron transmission cannot be ignored \cite{Galperin07}.
However, while rigorous computational treatments of 
vibrationally-coupled charge transport can be worked out
for small (few site) models, see e.g., Refs. \cite{Galperin07,Thoss09,Bijay15,Galperin20},  
in long junctions such methods reach their computational limit.
Alternatively, a common approximate approach to deal with the 
impact of nuclear effects on electronic conduction
is to employ classical molecular dynamics simulations for sampling molecular configurations, 
in conjunction with the Green's function formalism  for transport
\cite{Kubar09,Kubar1,Kubar2,Markussen17,Ignacio18,Ignacio19,Ignacio21}.  
Yet, even this approximate method becomes computationally impractical for large systems 
and it requires an additional level of coarse-graining \cite{Kubar1}. 
Moreover, this tool obviously fails to capture dynamical (contrasting static) effects 
arising due to the interaction of electrons with e.g., intramolecular nuclear vibrations and the solvent.

In this work, we study long-range charge transport in molecular junctions, 
homogeneous and modular.
Our goal is to explore how intrinsic molecular electronic parameters and environmental-thermal
effects play together to support long-range transport.
Specifically, our goal is to build models 
where quantum-coherent and thermal effects 
coexist and combine to contribute to effective transport characteristics.

To incorporate electronic decoherence and elastic and inelastic electron scatterings in the junction, 
we employ the Landauer-B\"uttiker probe (LBP) simulation technique \cite{Buttiker1,Buttiker2}.
In this approach, ``probes" mimic the interactions between conduction electrons and other physical degrees of freedom.
%
The LBP method has been originally developed to describe the coherent-to-incoherent turnover of conduction
in mesoscopic devices \cite{Buttiker1,Buttiker2}. 
More recently, the LBP technique has been used to study thermally-assisted
charge transport at the nanoscale, through
organic \cite{Kilgour1,Korol4} and biological molecules \cite{Anantram1,Anantram2, Anantram3,Korol1,Korol2,Korol3,Kim1,Kim2,Sun1, Sun2,Sun3}.
Applications of the LBP method to one-dimensional molecular junctions include studies of 
their electrical conductance \cite{Kilgour1}, nonlinear (high voltage) 
characteristics \cite{Kilgour2,Kilgour3}, thermopower \cite{Korol3}, 
and thermal conductance \cite{Malay,Roya}. The spin selectivity 
effect was examined in chiral structures 
such as in double-stranded DNA \cite{Sun1} and helical proteins \cite{Sun2} using LBP simulations, albeit 
at zero temperature. More recent applications of the LBP method include investigations of the
conductance of a multiterminal DNA tetrahedron junction under decoherence effects \cite{Sun3}.
Beyond studies of short (1-10 units) molecular junctions,
there have been fundamental applications of the LBP method
to understand scaling relations in transport. This includes
simulations of carriers' transport in periodic 
and quasi-periodic lattices under dissipation \cite{Goold21,BijayLBP},
and the coherent (ballistic) to Ohmic (diffusive) crossover in charge \cite{Pastawski,Dhar07} and 
phonon transport \cite{Kiminori1,Kiminori2}.

Our interest in exploring possible mechanisms of long-range charge transport 
is motivated by studies on electrically-conductive protein nanowires, specifically  
the \textit{Geobacter (G.) Sulfurreducens}. 
This bacteria, found in soil, reduces sulfur,
but the main interest in this system concerns its electrical capabilities \cite{bond2003}: 
\textit{G. Sulfurreducens} has a pilus (tail) that can transport 
charge up to the micrometer scale \cite{lovley2019}.
This observation has sparked much interest in the community, with recent works
reporting high rates of heme-to-heme electron transfer in multiheme cytochromem 
\cite{Butt,BeratanPNAS}, while 
other experiments questioning the nanowire capabilities of the Geobacter pili \cite{Gu20,Gu20Rev}.
Multiple theories have been proposed to decipher the workings of long-range biological 
charge transport, see e.g. Refs. 
\cite{agam2020,Beratan,Amdursky}, and references therein.
It was argued that charge transport in protein nanowires
cannot be satisfactory described by either a metallic-like conduction, or a completely 
classical-diffusive mechanism.
Rather, it was argued that a mixed coherent-incoherent mechanism takes place, with charge delocalization
over ``islands"  \cite{Beratan}. 
This conduction process could be facilitated by the varying rigidity of the 
molecule \cite{agam2020, Amdursky}.


In our work, we attempt to understand the interlaced role of internal electronic parameters 
and environmental-thermal effects 
on long-range charge transport, particularly in modular systems with segments of varying rigidity.
Using simulations, we probe the following questions: 
(Q1) What are some of the unique mixed coherent-incoherent transport mechanisms that can 
take place in {\it short} chains?
(Q2) When do inelastic-dissipative effects assist long-range transport, and when do they hamper it?
(Q3) What is the impact of structuring the molecule, to include segments of alternating rigidity, on 
long-range charge transport?

Models of electron transfer offer an enormous richness in their transport characteristics;
in our work here we discuss selected classes of models, and aim to draw general lessons 
from simulations.
Our results include the observation of nontrivial, non-monotonous trends in transport---once inter-site electronic couplings are made strong,
the observation of a transition from thermally-assisted to thermally-hampered
transport in long homogeneous chains, and an exposition of 
the intricate role of alternating molecular rigidity on electronic conduction. 
 


The plan of the paper is as follows.
In Sec. \ref{sec:mech} we recount conventional charge transport mechanisms 
that were theoretically proposed and experimentally observed
in molecular electronic junctions, and post several questions that guide us in our study.
In Sec. \ref{sec:method}, we briefly discuss the LBP method that we use throughout this paper.
Simulations for short chains are presented in Sec. \ref{sec:short}. 
Homogeneous long chains are analyzed in Sec. \ref{sec:longH}.
Systems of alternating environmental effects are examined in Sec. \ref{sec:longA}.
We conclude in Sec. \ref{sec:summ}.


\section{Conventional mechanisms and open questions}
\label{sec:mech}

\subsection{Setup and conventions}

Our setup includes a molecule connecting two metal electrodes in a metal-molecule-metal configuration.
We focus on linear, quasi-one-dimensional chains, and study the electronic conduction of junctions. 
A central nontrivial effect in this regard is the interaction of conduction electrons with other 
degrees of freedom, specifically nuclear motion in the form of vibrations.
While this many-body problem can be analyzed rigorously for small models 
such as the Anderson-Holstein model, the problem is very difficult to tackle
in a numerically-exact manner when many sites and vibrations are involved.

We now clarify our working assumptions in this study:
We only focus on low bias-voltage situations with the charge current being linear in the applied voltage bias. 
As such, the electrical conductance $G$ (or its inverse, the resistance $R$) 
is the relevant measure for charge transport.
Assuming non-magnetic components, we ignore the electron spin degree of freedom 
since it simply multiplies the electrical conductance by a factor of two.
We set the Fermi energy $\epsilon_F$ at zero; electronic site energies are measured relative to the 
Fermi energy.
As for the temperature, we assume that it is uniform
throughout the structure and identical for all degrees of freedom, nuclear and electronic.
We focus in this study on charge current only, but control of exciton dynamics
and realizing long-range exciton energy currents are similarly of an interest, 
and they display rich coherent-incoherent effects \cite{Aspuru17,Plenio,BeratanE}.

We refer to ``environmental effects" as factors leading to decoherence (phase loss) 
and elastic and inelastic scatterings of electrons. 
In the context of molecular junctions, these environmental effects arise from the 
motion of atoms in the molecule and the surrounding environments, e.g., the solvent.
The impact of environmental effects on electron transport should become more pronounced 
as we raise the temperature.
We note that in the LBP method, the influence of the environment is controlled by
scattering rate constants, $\gamma$. In reality, elastic and inelastic 
scattering effects depend on the temperature, as well as on the local environment, and
one could enrich the model to account for those aspects.

We focus on molecular junctions operating at temperatures in the range of 100-400 K, 
where many experiments are performed. 
We distinguish between short and long chains:
We refer to short chains as systems with $N=1-10$ units, roughly translating to
$0.5-5$ $nm$, assuming a unit (monomer) $\approx$ 5 \AA\ long, see e.g. Ref. \cite{Frisbie1,Frisbie2,Tao}.
In these junctions, one often observes the traditional mechanisms of tunneling, ballistic transport, 
and thermally-assisted hopping. 
Chains with $N= 20-80$ units (longer than $\approx$ 10 $nm$ but shorted than 50 $nm$) 
are considered ``long" here \cite{Barton34, Danny1, Danny2}, and the question of what 
transport mechanisms they display remains theoretically 
largely unexplored---and is a central focus of this work.
Longer chains with hundreds of sites, $\gtrsim$ 200 $nm$, approach the thermodynamic limit, 
and while interesting for understanding different phases in many-body physics 
(see e.g. Refs. \cite{Goold21,BijayLBP}), they are not considered in this
study. 

\subsection{Standard transport mechanisms in short chains}
\label{sec:mechS}

\begin{figure}[htbp]
\centering
\includegraphics[width=0.5\textwidth] {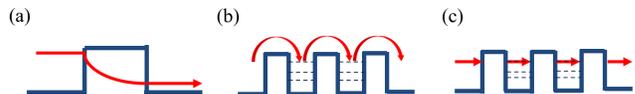} 
\caption{Schemes of standard transport mechanisms.
(a) In tunneling transmission, incoming electrons with kinetic energy below the barrier edge tunnel through.
(b) In the hopping mechanism, electrons occupy the molecular bridge and they hop through, 
activated by the thermal environment. 
(c) Ballistic transport concerns  electrons arriving with an energy matching 
molecular electronic levels, thus allowing resonant band-like transport.
In this scheme, the potential wells represent the molecular motifs with local electronic levels (dashed lines).
Arrows represent charge transport from left to right.
}
\label{fig:mech_types}
\end{figure}

In short metal-molecule-metal junctions of $N=1-10$ sites,
three main mechanisms typically show up, as sketched in Fig. \ref{fig:mech_types}:
(i) deep tunnelling, (ii) thermally-assisted hopping, and (iii) ballistic motion \cite{NitzanB,Cuevas2010}.  
The electrical conductance of many short junctions is 
adequately described using these classes, which we now elaborate on. 

In the tunnelling mechanism, the energy of molecular orbitals relevant for transport, 
e.g., the highest occupied molecular orbital (HOMO) is off-resonant from the Fermi energy.
This leads to electrical resistance that increases exponentially with the barrier  (molecular) size.
This mechanism dominates transport in short molecules of approximately up to 1 $nm$, 
such as the Benzenedithiol molecular junctions.

In longer molecular junctions (up to 10-15 units), 
thermally-assisted hopping or ballistic motion dominate the conductance, 
depending on the thermal and environmental conditions. 
In ballistic motion, electrons arriving from the metal, 
which are in resonance with molecular electronic orbitals, dominate transport.
As a result, this contribution, while susceptible to
scattering effects, is independent of molecular size. 
In contrast, in thermally-assisted hopping nuclear effects {\it promote} 
the thermal activation of electrons to occupy the molecule, 
as well as the hopping motion of electrons between sites.
In short chains of 1-10 sites, thermal effects enhance the total current, 
compared to the quantum-coherent frozen case by opening the hopping transport pathway.

\raggedbottom
\raggedbottom

The resistance $R$ of molecular junctions in these three different conventional cases is given by
the approximate scaling relations,
\raggedbottom
\bea
R\propto
\begin{cases}
e^{\kappa N} & \text{tunneling}\\
N \quad  & \text{hopping}\\
N^{0} & \text{ballistic}
\end{cases}
\eea
\raggedbottom
with $N$ the number of sites in the chain and $\kappa^{-1}$ a decay length 
characterizing the height and width of the tunneling barrier.

Note that the discussion of vibrationally-assisted mechanisms here is not intended to be comprehensive; other
transport mechanisms relying on static and dynamical fluctuations, and important for both
short and long-range transport, are e.g. through 
flickering resonances \cite{BeratanF},  quantum unfurling \cite{PeskinU},
variable-range hopping \cite{VRH}, and  polaron formation \cite{Polaron}.

\subsection{Open questions: Short and long-range charge transport}
\label{sec:Q}

We now pose the following questions, which concern mixed 
coherent and incoherent effects in short and long-range transport, and
direct us in this study:

(Q1) Beyond conventional mechanisms controlling charge transport in short chains,
as discussed in Sec. \ref{sec:mechS}, 
what other unique, mixed coherent-incoherent scenarios can develop? 
Specifically, in thermally-assisted transport,  (i) resistance  grows monotonically with size,
and (ii) conductance is enhanced with increasing thermal effects.
How can we tune the structure to display e.g., a
non-monotonic length dependence of the resistance?

(Q2) 
In short molecular junctions, the interaction of electrons with the environment leads to 
thermally-assisted hopping, which is beneficial for conduction.
However, in the classical diffusion limit, the resistance grows with length.
When does this crossover,
between thermally-assisted and thermally-suppressed conduction,  take 
place? What are the signatures of this crossover?

(Q3) Considering chains with alternating rigidity, e.g., made of 
segments of different electronic tunneling energies and 
environmental effects: What is the impact of this structuring on charge delocalization, and long-range 
charge transport?

In the next sections, we address these questions with minimal models and using LBP simulations. 
Given the richness of the problem, 
we do not aim for a comprehensive study of mixed quantum-classical transport
effects. Rather, we focus on specific classes of chains: homogeneous and modular.
Other systems leading to unconventional effects were investigated e.g. in Refs. \cite{Taomix1,Taomix2,Kim1}. 

\section{Model and Method}
\label{sec:method}

\subsection{Model Hamiltonian}
The molecular junction is described by the Hamiltonian
\bea
  \hat{H}=\hat{H}_W+\hat{H}_L+\hat{H}_R+\hat{V}_T+\hat{H}_P+\hat{V}_P.
  \label{eq:hamilton}
\eea
Here, $\hat{H}_W$ is the electronic Hamiltonian of the molecular chain.
$\hat{H}_L$, $\hat{H}_R$ are the Hamiltonians of the two metals, including collections of noninteracting electrons
with $\hat{V}_T$ as the tunneling Hamiltonian between the metals and the molecule.
$\hat{H}_P$ and $\hat{V}_P$ describe the probe contribution to the Hamiltonian, which are the source for decoherence and inelastic scatterings effects.
The molecule is described with  a tight-binding Hamiltonian with $N$ sites,
\bea
    \hat{H}_W = \sum_{n=1}^N\epsilon_n\hat{c}_n^{\dagger}\hat{c}_n+\sum_{n=1}^{N-1}\tau_{n,n+1}\hat{c}_n^{\dagger}\hat{c}_{n+1}+h.c.
    \label{eq:ham wire}
\eea
Here, $n=1,2,....,N$ is the site index. Sites do not necessarily correspond to a single atom:
a site could represent a monomer in organic polymers or a base-pair in a double-stranded DNA.  $\hat{c}_n^{\dagger}(\hat{c}_n)$ are fermionic creation (annihilation) operators to generate an electron at site $n$.
The parameters $\epsilon_n$ and $\tn$ are the site energies and the inter-site tunneling energies, respectively. 
We assume nearest-neighbor tunneling only. $h.c.$ stands for the hermitian conjugate.

The metal electrodes are similarly described with fermionic creation (annihilation) operators, $\hat{a}_{v,k}^{\dagger}(\hat{a}_{v,k})$, 
of momentum $k$ in lead $v$, 
\bea
    \hat{H}_v = \sum_k \epsilon_{v,k}\hat{a}_{v,k}^{\dagger}\hat{a}_{v,k}, \quad v=L,R.
    \label{eq:ham lead}
\eea
The tunneling Hamiltonian couples the
$L$ lead to site $1$  and the $R$ lead to site $N$ with tunneling energies $g_{v,k}$,
\bea
    \hat{V}_T = \sum_k g_{L,k}\hat{a}_{L,k}^{\dagger}\hat{c}_{1}+\sum_k g_{R,k}\hat{a}_{R,k}^{\dagger}\hat{c}_{N}+h.c.
    \label{eq:ham tun}
\eea
The essence of the LBP method is that the physical 
system is coupled to so-called ``probes", where decoherence and incoherent effects
are implemented. 
Mathematically and computationally, probes are described 
similarly to physical electrodes (albeit with additional constraints, see below).
As a result, the LBP method becomes tractable numerically and of a significant utility.

\begin{figure} [htbp]
\includegraphics[width = .35\textwidth]{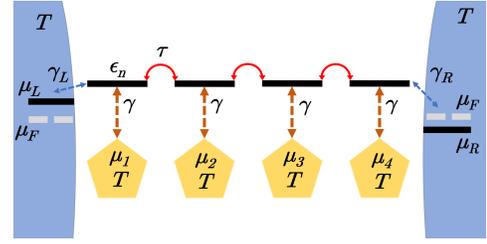} 
\caption{
Schematic diagram of a molecular junction, using probes to emulate environmental effects.
Molecular electronic sites are represented by full lines with 
energies $\epsilon_n$ and nearest-neighbor tunneling
$\tau$. Each molecular site suffers local thermal scattering effects emulated by a probe (pentagon)
with a rate constant $\gamma$.
Probes are metal objects, thus characterized by the Fermi distribution function 
given at temperature $T$ and chemical potential $\mu_n$ (solved for with the ``probe condition", 
Eq. (\ref{eq:linear})).  
The physical metals at the boundaries are coupled to the molecule with the hybridization energies 
$\gamma_{L,R}$.
}
\label{fig:Uni}
\end{figure}

We assume that each site $n$ is coupled to a local probe $n$, 
representing an independent local environment,
\bea
    \hat{H}_P = \sum_{n=1}^N\sum_k \epsilon_{n,k}\hat{a}_{n,k}^{\dagger}\hat{a}_{n,k}.
    \label{eq:ham probe}
\eea
Here, $\hat a_{n,k}^{\dagger}$ is a fermionic creation operator with momentum $k$ in the $n$th probe.
The tunneling between the molecule  and the probe is given by
\bea
    \hat{V}_P =  \sum_{n=1}^N\sum_k g_{n,k}\hat{a}_{n,k}^{\dagger}\hat{c}_{n}+h.c.,
    \label{eq:ham tran probe}
\eea
with $g_{n,k}$ as the tunneling energy.
The coupling of the metals (physical or probes) to the molecule is captured by hybridization functions,
\bea
    \label{eq: g to gam}
    \gamma_\alpha(\epsilon)=2\pi\sum_k|g_{\alpha,k}|\delta(\epsilon-\epsilon_{\alpha,k}),
\eea
Here, $\alpha$ is an index collecting the probes as well as the physical electrodes, $\alpha=1,2,...,N,L,R$ .
For simplicity, in simulations we assume the wideband limit and 
take $\gamma_\alpha$ as an energy-independent parameter.
We assume that the molecule is coupled with the same strength to the two metals, $\gamma_v=\gamma_{L,R}$.
The probe coupling parameter $\gamma_n$ controls the 
strength of interactions between the environment and conduction electrons. 
When $\gamma_n=0$ at every site, transport is  purely quantum-coherent and the conductance reduces to 
the Landauer formula.
Fig. \ref{fig:Uni} shows a scheme of the model 
with $\epsilon$, $\tau$, and $\gamma$ identical along the chain, representing a homogeneous system.

\subsection{The Landauer-B\"uttiker Probe method}

The LBP method was recently implemented for molecular junctions in Ref. \cite{Korol4}. 
For completeness, we review it below.
The index $\upsilon$ identifies the physical metal electrodes at the boundaries, $L$,$R$; 
$n=1,2,..,N$  stands for the probe terminals. 
The index $\alpha$ is used to identify both physical leads and probes.
The net charge current flowing in the system is calculated at the left site,
\bea
    \label{eq:current}
    I_L=\frac{e}{2\pi\hbar} \sum_\alpha \int_{-\infty}^{\infty}\mathcal{T}_{L,\alpha}(\epsilon)\left[f_L(\epsilon)-f_\alpha(\epsilon)\right]d\epsilon,
\eea
where $f_\alpha(\epsilon)=\left[1+\exp(\beta(\epsilon-\mu_\alpha))\right]^{-1}$
is the Fermi function with $k_bT=\beta^{-1}$ and  
$\mu_\alpha$  the chemical potential of the $\alpha$ terminal. 
The chemical potentials of the probes are calculated from the particle conservation condition,  Eq. (\ref{eq:linear}), which we describe below.
Eq. (\ref{eq:current}) is given in terms of the
transmission function 
\bea
\mathcal{T}_{\alpha,\alpha'}(\epsilon)=\trace\left[\Hat{\Gamma}_\alpha(\epsilon) \Hat{G}^r(\epsilon)\Hat{\Gamma}_{\alpha'}(\epsilon)\Hat{G}^{a}(\epsilon)\right],
 \label{eq:transmission}
\eea
with $\Hat{G}$ being the retarded Green's function, 
$\Hat{G}^r=\left[\Hat{I}\epsilon-\Hat{H_W}+i\sum_{\alpha}\Hat{\Gamma_{\alpha}}/2\right]^{-1}$
and $[\Hat{G}^r]^{\dagger}=\Hat{G}^a$. $\Gamma_\alpha$ are the hybridization matrices.
In our setup, all elements in these matrices 
are null except in one place, corresponding to the site that is connected to a physical metal or a probe,
\bea
    \begin{split}
        &[\hat{\Gamma}_n(\epsilon)]_{n,n}=\gamma_n, \quad n=1,2,...,N\\
        &[\hat{\Gamma}_L(\epsilon)]_{1,1}=\gamma_L, \quad [\hat{\Gamma}_R(\epsilon)]_{N,N}=\gamma_R.\\
    \end{split}
    \label{eq:gammas}
\eea
Recall that we work in the wideband limit.
Having built the transmission function, we solve for 
the probes' chemical potentials at each site by enforcing current conservation on the physical system: 
The net charge current flowing between each probe and the system must be zero,
\bea
    \label{eq:current_zero}
    I_n=\frac{e}{2\pi\hbar} \sum_\alpha \int_{-\infty}^{\infty}\mathcal{T}_{L,\alpha}(\epsilon)\left[f_n(\epsilon)-f_\alpha(\epsilon)\right]d\epsilon=0. 
\eea
Under low applied bias, that is, in linear response, this condition translates 
to a set of $N$ linear equations,
\bea
    \label{eq:linear}
    \nonumber
    &&\mu_n\sum_\alpha\int^\infty_{-\infty}\left(-\frac{\partial f}{\partial\epsilon}\right)\mathcal{T}_{n,\alpha}(\epsilon)d\epsilon\\
    &&\nonumber
    -\sum_{n'}\mu_{n'}\int^\infty_{-\infty}\left(-\frac{\partial f}{\partial\epsilon}\right)\mathcal{T}_{n,n'}(\epsilon)d\epsilon\\
    &&
    =\int^\infty_{-\infty}d\epsilon\left(-\frac{\partial f}{\partial\epsilon}\right)\left[\mathcal{T}_{n,L}(\epsilon)\mu_L+\mathcal{T}_{n,R}(\epsilon)\mu_R\right],
\eea
which can be written in a compact form as $M\mu=v$. 
Solving this system yields the chemical potentials of each probe, $\mu_n$, used in turn
in Eq. (\ref{eq:current}) to calculate the net low-bias current,
\bea
I = \frac{e}{2\pi\hbar}\sum_\alpha\left[\int_{-\infty}^\infty\mathcal{T}_{L,\alpha}(\epsilon)\left(-\frac{\partial f}{\partial\epsilon}\right)d\epsilon\right](\mu_L-\mu_\alpha).
\nonumber\\
\eea
Finally, the electrical conductance is given by $G\equiv I/\Delta V$, and its inverse $R=1/G$ is the resistance of the junction.


\section{Short homogeneous chains}
\label{sec:short}

We begin our analysis with Q1 (Sec. \ref{sec:Q}) on realizing unconventional transport trends.
We focus first on short homogeneous chains with $N=1-10$ identical units and assume that all sites experience 
the same degree of environmental decoherence and scatterings,
$\gamma=\gamma_n$. To clarify trends, we typically extend molecules to up to 15 sites.
In simulations, we discretize integrals assuming electronic bands with hard cutoffs at $D=\pm 5$ eV, 
approximating the wideband limit. The voltage is assumed small 
with $\mu_L=-\mu_R=0.005$ eV. 

\begin{figure}[htbp]
\hspace{-2mm}\includegraphics[width = .5\textwidth]{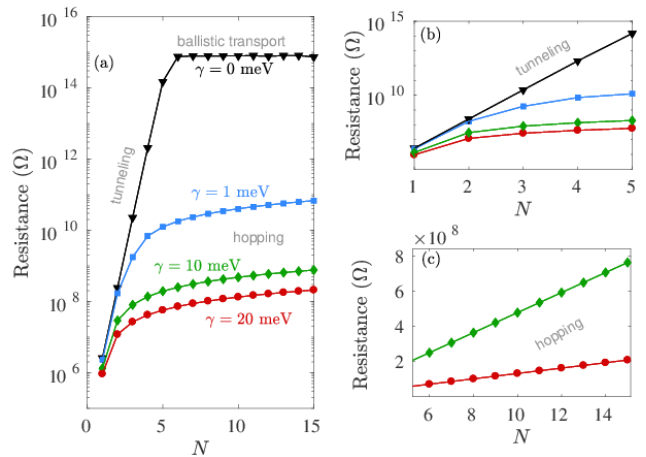} 
\caption{Electrical resistance as a function of molecular size
for short chains with $N=1-15$ sites at
 small intersite tunneling energies, $\epsilon_n\gg \tau$.
(a) Three different mechanisms: deep tunneling, ballistic motion and thermally-assisted hopping
are exposed when inspecting the resistance as a function of size and $\gamma$.
(b) Coherent, deep-tunneling behavior is manifested for short chains with an exponential
enhancement of resistance  with size ($\bigtriangledown$).
(c) Hopping behavior develops as we increase $\gamma$ in long systems (note the linear scale).
Parameters are $T=200$ K, $\epsilon_n=0.5$ eV, $\tau=0.05$ eV and $\gamma_{L,R}=0.05$ eV. }
\label{fig:multi_mich}
\end{figure}
%
\begin{figure}[htbp]
\includegraphics[width = .48\textwidth]{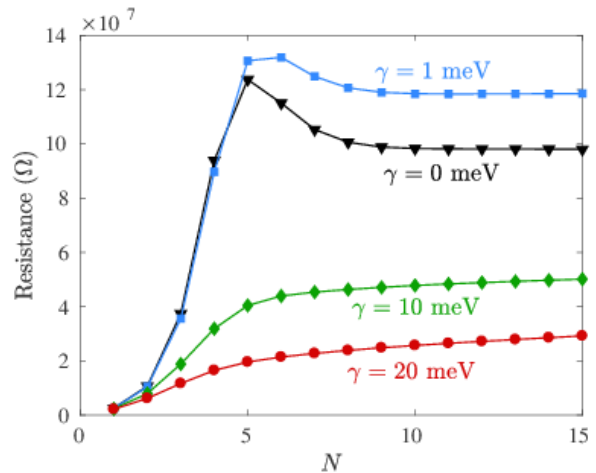} 
\caption{Electrical resistance as a function of molecular size for short chains with
large tunneling energies, $\epsilon_n\approx \tau$. 
Parameters are the same as in Fig. \ref{fig:multi_mich}, besides $\tau=0.2$ eV.}
\label{fig:multi_me}
\end{figure}

We first study in Fig. \ref{fig:multi_mich} a setup that displays what we regard as conventional transport.
Here, the intersite tunnelling energies are taken much smaller
than site energies (taken relative to the Fermi energy of the metals), $\tau/(\epsilon_n-\epsilon_F)\ll1$.
In accordance with theoretical and experimental studies, see e.g. Refs. \cite{NitzanB,Frisbie2,Kilgour1},
this junction demonstrates a smooth transition from the tunnelling regime in short molecules  
to the thermally-assisted hopping mechanism for $N\gtrsim 3$, once environmental effects are active 
with $\gamma\neq0$.
A transition from deep-tunneling to ballistic motion shows up when $\gamma=0$.
Organizing our observations, we recount the following trends in Fig. \ref{fig:multi_mich}:

(i) Increasing the probe coupling $\gamma$ lowers the resistance due to the contribution of the 
hopping mechanism. In Ref. \cite{Kilgour1}, we showed that at large enough $\gamma$ (order of few eVs,
unphysical in the context of molecular junctions) a Kramers-like turnover \cite{Hanggi}
takes place and the resistance begins to increase with $\gamma$.
(ii) Hopping resistance increases linearly with size, 
 which can be seen in Fig. \ref{fig:multi_mich}(c). 
(iii) Tunnelling is demonstrated for very short chains. 
This mechanism becomes less dominant 
as the system size increases: In isolated systems ($\gamma=0$), 
the resistance becomes length-independent in long enough chains,
indicative of the ballistic regime.  In contrast, under environmental effects 
the resistance becomes Ohmic with $R \propto N$.

Contrasting this conventional scenario, in Fig. \ref{fig:multi_me} we increase the 
electronic intersite tunneling energy,
making it now comparable to site energies, $\tau\approx (\epsilon_n-\epsilon_F)$.
Compared to  Fig. \ref{fig:multi_mich},
this situation represents a more rigid molecular structure, which better facilitates
tunneling between sites.
Fig. \ref{fig:multi_me} shows two nontrivial results:
(i) The resistance is no longer monotonically increasing with $\gamma$. 
In fact, the resistance is higher by about 20\% when $\gamma=1$ meV, compared to the isolated case.
(ii) The resistance no longer increases monotonically with size. 
Instead, a local maxima shows for $N\sim 5$ when $\gamma$ is small.

\begin{figure}[htbp]
\centering
\includegraphics[width=0.5\textwidth]{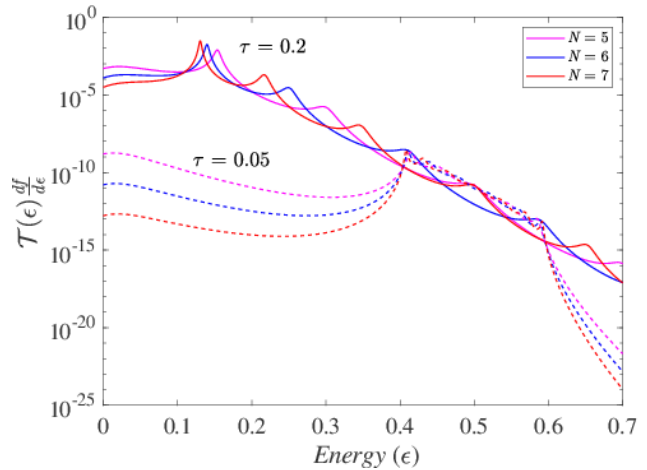} 
\caption{Transmission function multiplied by $\frac{df}{d\epsilon}$ and plotted as a function of the energy of incoming electrons.
Simulation parameters are  $\epsilon_n=0.5$ eV, $T=200$ K, $\gamma = 0$
for (full) $\tau=0.2$ eV and (dashed) $\tau=0.05$ eV, with three molecules of different sizes.
    }
    \label{fig:trans_times}
\end{figure}

Aiming to elucidate these phenomena, we study the isolated case of $\gamma=0$ in Fig. \ref{fig:trans_times}, 
by analyzing the energy-resolved conductance for $\tau=0.05$ eV and  $\tau=0.2$ eV, corresponding to Figs. \ref{fig:multi_mich} and \ref{fig:multi_me}, respectively.
Recall that in the coherent case ($\gamma=0$), the electrical conductance is proportional 
to the integral of the transmission function multiplied by 
the derivative of the Fermi function. 
There are two main contributions to the conductance integral: 
(i) Near the Fermi energy, $\epsilon=0$. 
(ii) Around energies of molecular orbitals, $\epsilon\approx\epsilon_n\pm \tau$. 

According to Fig. \ref{fig:trans_times},
for small $\tau$ and in short chains 
the dominant contribution to the conductance arises from the region near the 
Fermi energy (scenario (i)). However, as more sites are added, the major contribution shifts to
the transmission peaks (scenario (ii)). This latter contribution hardly changes with
size and it leads to the ballistic transport with resonant thermal electrons dominating transport.
In contrast, when $\tau\approx\epsilon_n$, the transmission functions are broad with
resonant features pushed closer to the Fermi energy, thus dominating the conductance (scenario ii).
As we increase the molecular size, the peak structure is further pushed towards the Fermi energy
thus enhancing (reducing) the conductance (resistance). 
This effect, of the suppression of resistance with increasing length,
is slowly diminishing with the introduction of each successive site, 
leading to the eventual plateau seen in Fig. \ref{fig:multi_me}, indicative of ballistic motion.

As to the non-monotonic behavior of the resistance with $\gamma$ at small $\gamma$, it cannot be easily  
examined by watching the transmission function since many probes contribute to the overall conductance. 
It remains a curious observation that merits further explorations.

Concluding this part, which focused on short homogeneous chains:
We showed that by increasing the intersite tunneling energy, representing a more rigid structure, 
the resistance of short molecules developed nontrivial trends. 
Specifically, 
in rigid systems with weak environmental effects the resistance no longer increased monotonically with size.


\section{Long homogeneous chains}
\label{sec:longH}

For short chains with $N=1-15$ sites, 
Figs. \ref{fig:multi_mich} and \ref{fig:multi_me} had taught us that environmental effects 
in the form of $\gamma\neq 0$ are generally {\it advantageous} for transport, 
activating the hopping conduction.
We now address Q2: What is the impact of environmental effects on transport in long chains? 
We expect that dissipation should eventually lead to an increase in the junction's resistance in long enough 
chains, compared to the ballistic limit. 
But how long should the chain be to demonstrate this environmental-hindered transport effect?

\begin{figure}[htbp]
\includegraphics[width =  .48\textwidth]{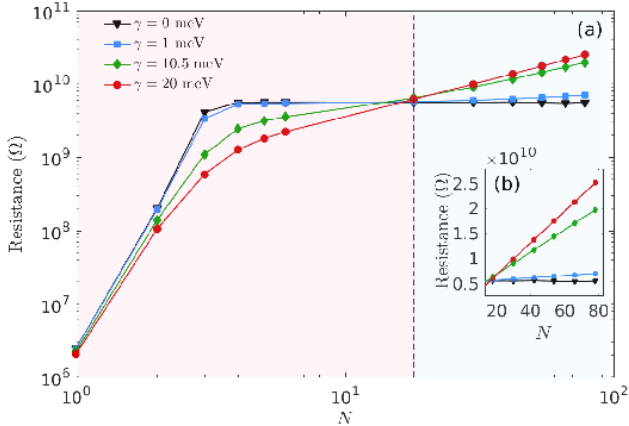} 
\caption{Long-range charge transport in homogeneous chains  as a function of size at different environmental
conditions, $\gamma$.
(a) Crossover from environmentally-assisted (pink, left of dashed line)
to environmentally-hindered (light blue, right of dashed line) transport.
(b) Zoom over the large-$N$ behavior demonstrating that $R\propto N$ for nonzero $\gamma$.
Parameters are the same as in Fig. \ref{fig:multi_mich},
$\epsilon_n = 0.5$ eV, $\tau=0.05$ eV, only at higher temperature of $T=400$ K.}
\label{fig:Uniform1}
\end{figure}

\begin{figure}[htbp]
\includegraphics[width =  .48\textwidth]{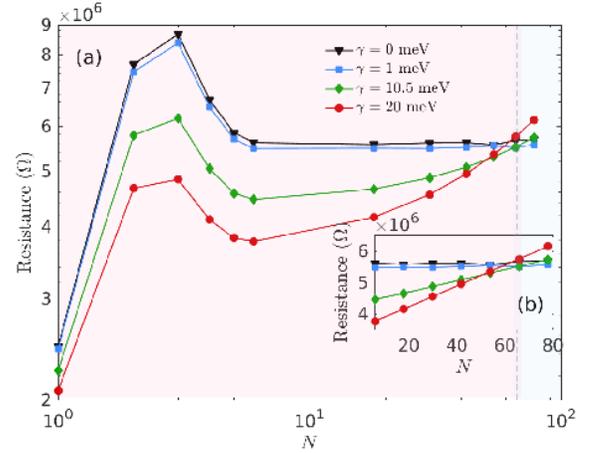} 
\caption{Long-range charge transport in homogeneous chains.
(a) Crossover from environmentally-assisted (pink) to
environmentally-hindered (light blue) transport, as revealed
in the behavior of the resistance as a function of size at different environmental conditions $\gamma$.
(b) Zoom over the large $N$ behavior, demonstrating that $R\propto N$ for large enough $\gamma$.
Parameters are the same as in Fig.  \ref{fig:multi_me},
specifically $\epsilon_n = 0.5$ eV, $\tau=0.2$ eV but $T=400$ K.}
\label{fig:Uniform2}
\end{figure}

Results of simulations are presented in Figs. \ref{fig:Uniform1} and \ref{fig:Uniform2}. 
Here, we study chains of up to $N\approx80$ sites
with varied probe values in the range of 1 to 20 meV, comparable to the thermal energy.
In analogy to short chains, we consider two cases, of weak ($\tau=0.05$ eV) 
and intermediate-strong ($\tau=0.2$ eV) intersite tunneling energies. 
Our simulations clearly show that while for short system (left region, pink) 
the environment assists transport and the resistance is reduced when increasing $\gamma$, 
long systems (right region in light blue) manifest the opposite trend; as we increase $\gamma$,
the resistance grows.
For completeness, we also display in Fig. \ref{fig:Uniform1} the short-system  behavior 
(similar to Fig. \ref{fig:multi_mich}, albeit at a higher temperature) and showcase the full crossover picture.
Similarly, Fig. \ref{fig:Uniform2} extends the range of Fig. \ref{fig:multi_me}.

Note that to exemplify this crossover---from 
environmentally-assisted to environmentally-hindered transport---we set the temperatures at $400$ K;
 at lower temperatures this transition take
 place for longer chains with hundreds of sites, where simulations become difficult to perform.
%
Comparing Figs. \ref{fig:Uniform1} to \ref{fig:Uniform2}, of
small and large intersite tunneling, respectively, we find that in the latter case 
the advantageous role of the environment on transport persists up to $N\approx60$. 
As well, it is worthwhile to highlight that the resistance for the large-$\tau$ junction is 
significantly lower compared to the small-$\tau$ case.

Overall, while in short chains transport is enhanced with increasing interaction to the surrounding environment, the opposite trend is observed in long chains  where the most effective transport is metallic-like (ballistic, non-dissipative).


\begin{figure}[htbp]
\includegraphics[width =  .48\textwidth]{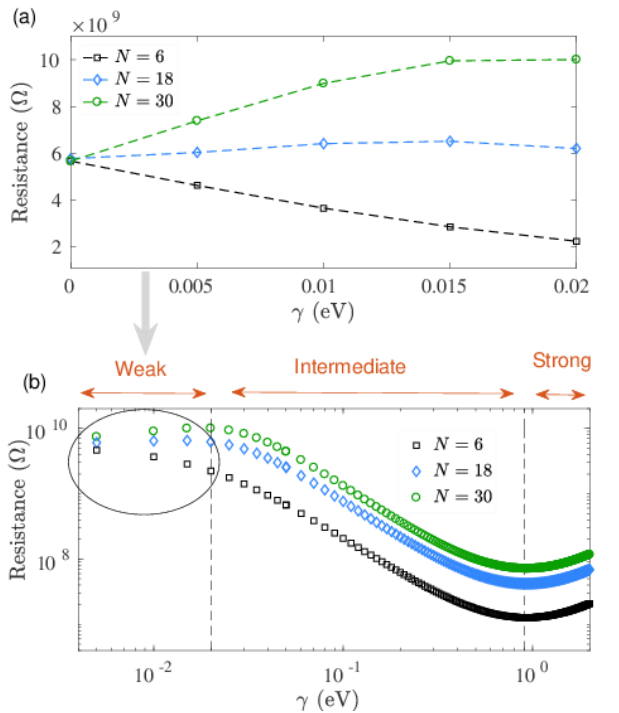} 
\caption{Resistance as a function coupling energy $\gamma$.
(a) The weak dissipation limit, which is relevant for this study, brings rich behavior with
environmentally-assisted transport for $N=6$ (square), environmentally-hindered transport for $N=30$ (circle),
 and a unique, environmentally-passive transport for $N=18$ (diagonal).
(b) Short chains ($N=6$, square) display a Kramers-like turnover behavior from
environmentally-assisted transport at weak-intermediate coupling
to the strong damping limit with transport hindered by $\gamma$.
The situation however is more rich for long systems ($N=18$, diagonal; $N=30$, circle).
Parameters are the same as in Fig. \ref{fig:Uniform1}. }
\label{fig:Fig5b}
\end{figure}

Complementing the length dependence investigation,
in Fig. \ref{fig:Fig5b} we study how the resistance depends on the environmental coupling $\gamma$.
In  Fig. \ref{fig:Fig5b}(a), we focus on the  range $\gamma = $ 1-20 meV, 
comparable to the thermal energy and relevant for molecular conductors.
In line with the discussion above, short chains benefit from enhanced $\gamma$ while the resistance of 
long chains of e.g. $N=30$ grows with $\gamma$. 
In between, we find that 
the resistance of a ``magic" chain with $N=18$ sites is robust against the environment. This is 
a unique effect where the negative impact of the environment compensates its advantage, 
leading to $R$ being about fixed with $\gamma$.
The specific size at which this novel effect shows up depends 
on the energy parameters of the chain and the temperature.

As we further raise $\gamma$ from 50 meV to 1 eV (values that are irrelevant for molecular conductors) 
we observe nontrivial trends in Fig. \ref{fig:Fig5b}(b) with the resistance behaving non-monotonically. 
We roughly distinguish in panel (b) between the weak and intermediate dissipation limits 
and the strong dissipation limit, in accord with Kramers' turnover behavior. 
However, unlike the original Kramers' model, 
the behavior in  Fig. \ref{fig:Fig5b}(b) is quite complex since it further depends on the molecular length.

\begin{figure} [htbp]
\includegraphics[width = .48\textwidth]{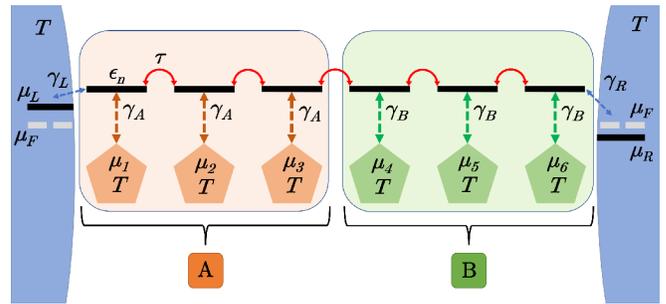} 
\caption{
Schematic diagram of the alternating-block motif. In this modular chain, 
three sites share the same probe coupling conditions ($\gamma_A$),
followed by three sites of a different probe coupling condition ($\gamma_B$).
Such units of 6 total sites are repeated in our simulations.}
\label{block_sys}
\end{figure}

\section{Modular chains}
\label{sec:longA}

We turn to  Q3 (see Sec. \ref{sec:mech}) probing
the impact of structuring the chain on long-range transport.
We limit the discussion to specific type of inhomogeneity, focusing on 
structures that include segments of alternating molecular rigidity.

Long-range charge transport has been reported in the \textit{G. Sulfurreducens} 
bacteria up to a micrometer scale \cite{agam2020}. 
Studies have shown that the geobacteria's long-range transport cannot be replicated by solely 
using hopping or coherent transport mechanisms \cite{Beratan}. 
Newly-proposed mechanisms have been suggested based on the alternating rigidity of the protein conducting the 
charge in biological nanowires \cite{agam2020,Amdursky}.

Inspired by biological nanowires, 
in this section, we explore the use of probe ``blocks" to mimic modular nanowires of alternating 
rigidity and study the resulting effect on the resistance of the system. 
We begin by assuming that the electronic structure is uniform, but that there are segments 
of the molecule that are better protected from the environment, and thus are more rigid (frozen). 
Other segments  are prone to environmental effects, with electron transport
there interacting more strongly with intramolecular or intermolecular degrees of freedom.
We hypothesize that such a structured system could support mixed coherent-incoherent transport,
with rigid blocks supporting delocalized charge transport, and ``flexible" sections supporting Ohmic conduction.

To address this type of system, we build a chain with blocks of alternating probe condition,
see Fig. \ref{block_sys}.
For computational simplicity, each block is made of three sites. 
For a given block, we set a probe coupling parameter at $\gamma_A$, 
followed by three sites with a different coupling parameter, $\gamma_B$.


\subsection{Resistance}

\begin{figure}[htbp]
\includegraphics[width = .48\textwidth]{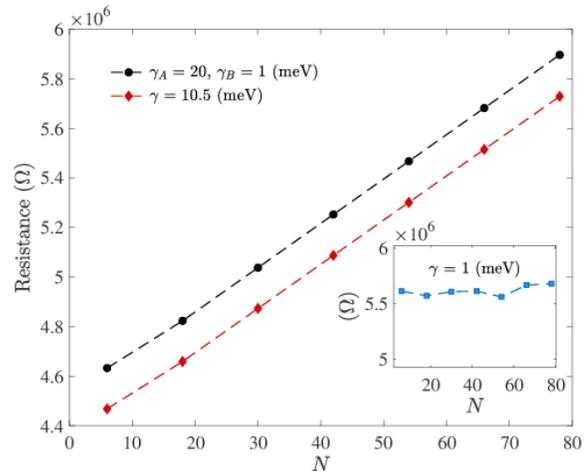} 
\caption{Modular chains.
Resistance as a function of chain size 
plotted for homogeneous systems ($\gamma=10.5$ meV, diagonal) and alternating-$\gamma$ systems
($\gamma_A=20$, $\gamma_B$=1 meV, circle). 
While the resistance per site is nearly identical in the two cases,
the block system overall has a higher resistance. 
The inset shows results for the uniform chain with $\gamma=1$ meV, demonstrating ballistic transport.  
Parameters are $T=400$ K, $\epsilon_n=0.5$, $\tau=0.2$, $\gamma_{L,R}=0.05$ (eV).}
\label{mix_ht}
\end{figure}

In Fig. \ref{mix_ht}, we show simulation results for the resistance, jumping in 
units of six sites (AB block).
In addition to simulating an alternating system, 
we also present results for a uniform ``averaged" chain with
$\gamma=(\gamma_A+\gamma_B)/2$.

\begin{figure*}[htb]
\includegraphics[width = .45\textwidth]{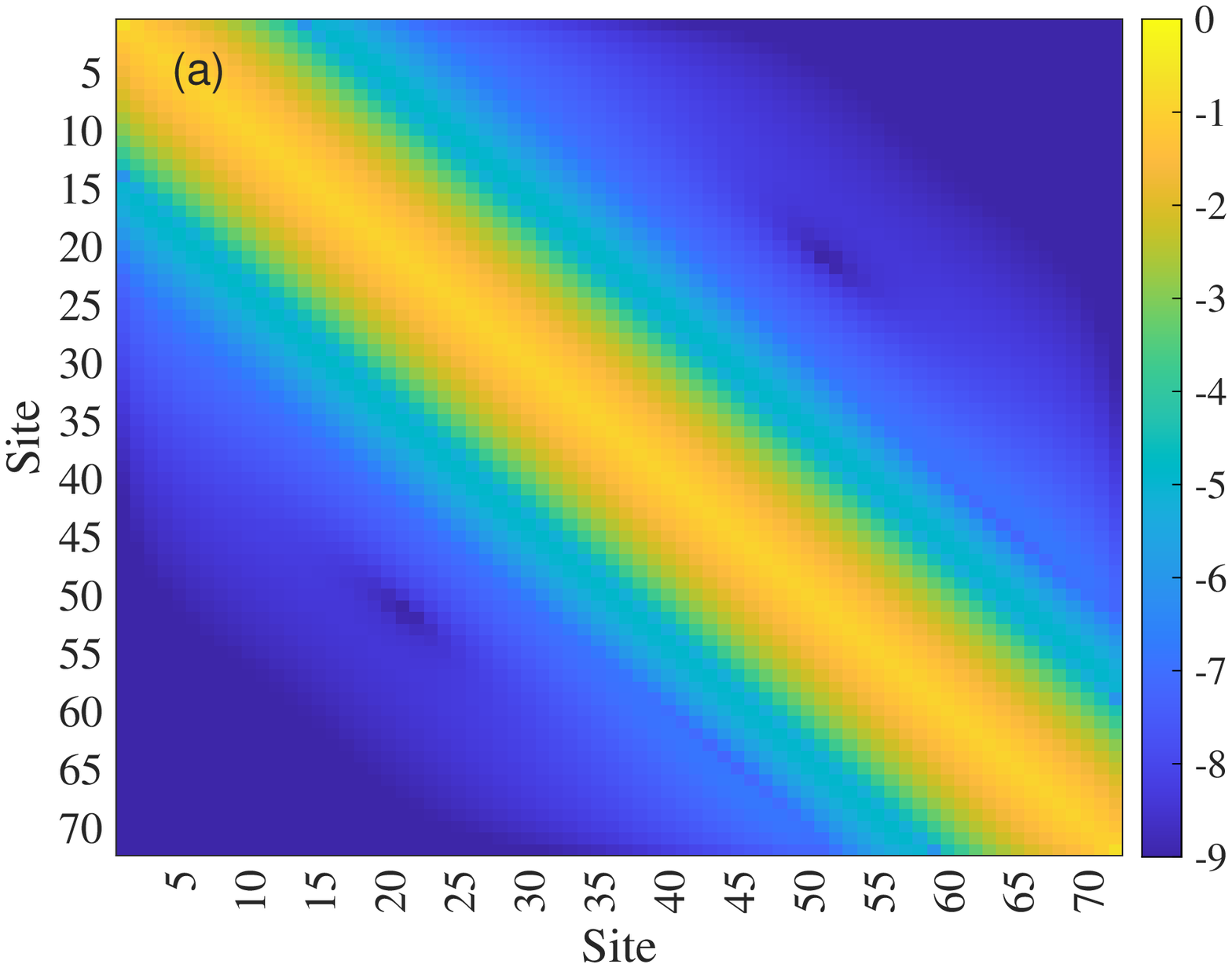} 
\includegraphics[width = .45\textwidth]{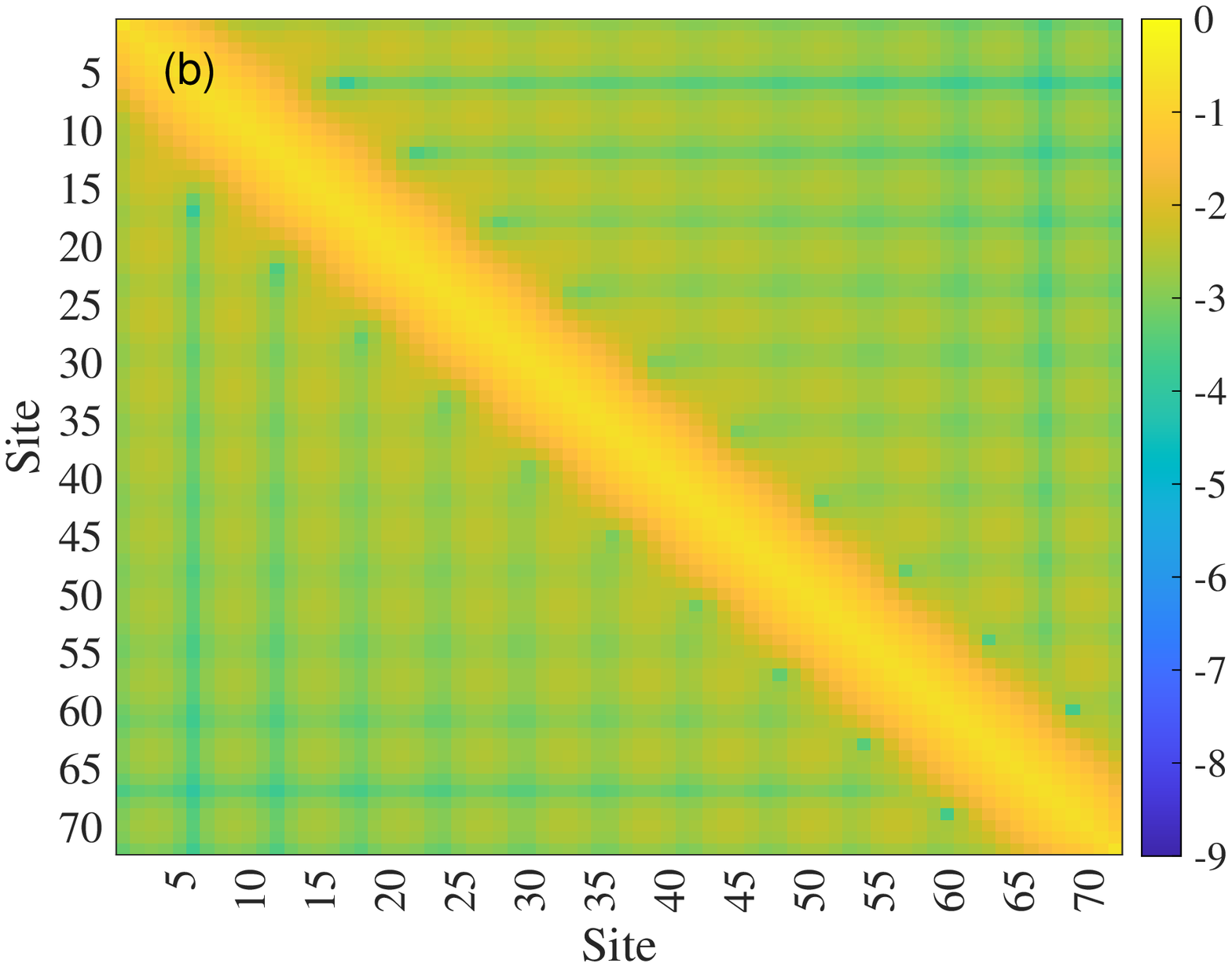} 
\caption{Charge correlation analysis.
Density matrix of (a) a homogeneous chain ($\gamma=10.5$ meV) and (b) an alternating  system
($\gamma_A=20$, $\gamma_B=1$ meV).
For clarity, we display  $\log_{10}|\rho_{n,m}|$.
Correlations in the alternating system are more significant
and extended than in the uniform system.
Parameters are  $\epsilon_n = 0.5$ eV, $\tau=0.2$ eV, $N=72$;
other parameters are the same as in Fig. \ref{mix_ht}.}
\label{fig:avg_den}
\end{figure*}


One important observation from Fig. \ref{mix_ht} is that the resistance of the 
alternating chain is simply Ohmic, $R\propto N$. This result is nontrivial: A homogeneous chain with
$\gamma=1$ meV supports ballistic motion (inset), 
while a long homogeneous chain with $\gamma=20$ meV leads to an Ohmic behavior (Fig. \ref{fig:Uniform2}).
Naively, one would expect that a chain alternating between those two limits
would manifest  mixed characteristics of these two mechanisms.
Instead, we find that the alternating junction shows a simple Ohmic behavior, and with 
the {\it same slope as the homogeneous chain of an averaged coupling}, $\gamma=(\gamma_A+\gamma_B)/2$.
The difference between the resistance of the two systems shown in Fig. \ref{mix_ht} probably
corresponds to the contact resistance, which is distinct in the alternating and the averaged cases.
In addition to the AB alternating system, we also simulated a BA alternating system, 
which yielded the same results. 

As an additional note, in studies of the \textit{G. Sulfurreducen}, the efficiency 
of long-range transport had been attributed to its alternating rigidity. 
In contrast, our alternating system resulted in a somewhat 
higher resistance compared to the averaged system. 


\subsection{Correlations}

To better understand the averaging behavior seen in Fig. \ref{mix_ht}, where the resistance per site of a modular system equals that in
an averaged-$\gamma$ homogeneous chain,
we inspect charge correlations in the alternating and uniform systems,
$\rho_{n,m}={\rm Tr}\left[\hat c_n^\dagger \hat c_m \rho \right]$, 
where $\hat c^\dagger_n$ and $\hat c_n$ are creation and annihilation operators of electrons on the 
molecular site $n$, respectively. 
We obtain the density matrix of the system using the 
Green's function approach  \cite{Cuevas2010},
\bea
\rho_{j,k}=\frac{1}{2\pi}\sum_\alpha\int^\infty_{-\infty} d\epsilon
\left[\hat G^r(\epsilon)\hat \Gamma_\alpha \hat G^\alpha(\epsilon)\right]_{j,k}f_\alpha(\epsilon).
\label{eq:dens_calc}
\eea
Recall that $\hat \Gamma_\alpha$ is a system-metal hybridization matrix and $f_\alpha(\epsilon)$ 
is the Fermi function of a probe or real metal. 
Site populations appear on the diagonal of the density matrix, while correlations between sites
appear on the off-diagonal elements.

Results are plotted in Fig. \ref{fig:avg_den}, and we study both homogeneous and modular-alternating systems.
Comparing the two chains we note the following:
(i) In the alternating case,   Fig. \ref{fig:avg_den}(b),
the density matrix has a grid-like pattern due to the alternating $\gamma$ setting. 
This grid spans about six sites, which corresponds to the size of the repeating unit.
In contrast, correlations in the homogeneous chain quickly decay away from the diagonal,  
Fig. \ref{fig:avg_den}(a).
(ii) Compared to the uniform system, the alternating chain shows significantly stronger correlations.
For example, the correlation $|\rho_{18,60}|$ is about 10$^{-8}$ in the homogeneous case, while in the 
 modular case,  $|\rho_{18,60}|\approx 10^{-3}$.
The intriguing  observation is that, although the alternating chain
has larger and more extended correlations than the homogeneous system, 
these internal correlations do not translate into  qualitatively-different resistances
in the two cases, see Fig. \ref{mix_ht}. 

Our motivation in introducing alternating-$\gamma$ chains has been to 
examine the role of alternating rigidity on the resistance. 
We had  envisioned that such structuring would allow long-range transport due to partial delocalization, with
higher currents than the averaged limit. 
However, our simulations show that this is not the case: 
The resistance of an alternating chain is about the same as of an 
homogeneous chain of an averaged $\gamma$. 
We found however that charges in the system are 
delocalized and correlated over many sites.

To further test these seemingly incompatible observations, we analyze a related problem in Appendix A.
There, we study the steady-state flux of particles through homogeneous  or alternating chains
 (without metal electrodes) by adopting the Lindblad quantum master equation (QME).
We find that, once again the flux (which is related to the electrical 
conductance in the LBP simulations)
is insensitive to internal modulations of $\gamma$, and it shows an averaging effect.
This qualitative agreement, between the Landauer B\"uttiker probe method and the Lindblad QME 
is significant, and it supports the physicality of the LBP simulations.


\subsection{Chains with alternating electronic and environmental parameters}

To break the averaging effect, it is clear that we need to further enrich the model, 
e.g., by structuring electronic parameters on top of the environmental patterning. We present here an example
of such a setup. We use again two different blocks, A and B, 
now characterized by alternating both electronic parameters $\tau$ 
and environmental coupling $\gamma$ using ($\tau_A$, $\gamma_A$) 
in block A and ($\tau_B$, $\gamma_B$) in B. 
By introducing these alternating parameters, one can more
 closely mimic biological nanowires of alternating rigidity. 
In the enriched model, we pair a small probe coupling value with large intersite tunneling energy, and vice-versa:
In rigid regions, the electronic coupling is large, and the system is less prone to environmental effects.
%
In addition to this system, we simulate a chain with the opposite pairings of 
$\gamma$ and $\tau$.

\begin{figure}[h]
\includegraphics[width = .48\textwidth]{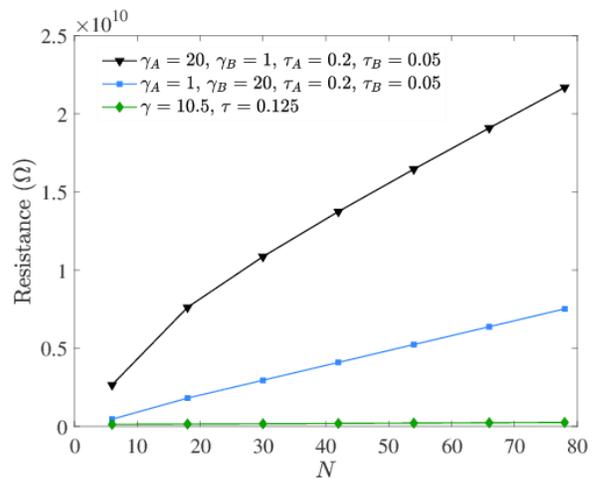} 
\caption{
Modular system consisting of alternating blocks with ($\gamma_A$, $\tau_A$) and ($\gamma_B$, $\tau_B$).
Simulations conditions are the same as in Fig. \ref{mix_ht},
except for $\tn$.} 
\label{fig:lbp_gamma}
\end{figure}

Results are presented in Fig. \ref{fig:lbp_gamma}. 
The blue line (square) corresponds to the physical pairing representing blocks of high and low rigidity. 
When the pairing is mixed, we get the black line (triangle). 
Indeed, with the modulation of both electronic structure parameters and 
environmental effects, we find a notable difference between the two situations both
in the total resistance, and the resistance per site, with the slopes differing by roughly a factor of 4. 
The resistance however is still of an Ohmic nature;
it remains a challenge to enrich the model such that charge transport would display unique characteristics---beyond Ohmic trends---in long chains.

With regard to Q3, we thus encountered several fundamental aspects, that (i) modular environmental effects are insufficient to imprint resistance distinct from the averaged value, unless the electronic parameters are further modified, and that (ii) Ohmic resistance may persist even when charge delocalization effect  is significant. 

\section{Conclusions}
\label{sec:summ}

We studied short and long-range charge transport in one-dimensional, homogeneous and modular molecular chains
looking for unconventional transport trends.
For short molecules, we showed that beyond 
 traditional trends associated with deep-tunneling, ballistic motion, and environmentally-assisted transport, 
junctions can show nonmonotonic characteristics (resistance vs. size) 
when the internal electronic coupling is increased.
In long homogeneous chains, we demonstrated the crossover from environmentally-assisted to 
environmentally-hindered transport as one increases the molecular length. 
We further showed that there are cases when parameters play together to 
generate length-independent resistance even when the system is coupled to the environment.


Inspired by biological electron transfer, we constructed an alternating-block system 
where some sections of the molecule were exposed to the environment, while other segments were protected.
While chains made of each different segment (A or B) showed distinct conductance (ballistic-delocalized 
vs. Ohmic), the modular AB system did not exhibit unique transport behavior, and
the resistance was simply Ohmic and of a similar magnitude to what one would get for 
an homogeneous chain with an A-B averaged probe parameter. 
However, the alternating system supported
stronger correlations in the chain, pointing to the fact that 
quantum delocalization effects were at play, though not manifested in the overall conductance.

In a broader context, our simulations show that by solely engineering dissipation on different sites 
one cannot control the resistance and it behaves as in  the averaged-homogeneous case.
In future work, we will explore the emergence of novel transport regimes 
under the control of both electronic and environmental conditions. 
Further benchmarking the LBP method against microscopic models is necessary, to validate the 
physicality of the model.



\begin{acknowledgments}
DS acknowledges the NSERC discovery grant and the Canada Research Chair Program.
\end{acknowledgments}

\vspace{5mm}
\noindent The data that support the findings of this study are available from the corresponding author upon reasonable request.


\renewcommand{\theequation}{A\arabic{equation}}
\setcounter{equation}{0}
\setcounter{section}{0} 
\section*{Appendix: Long-range charge transport with the Lindblad Quantum Master Equation}
\label{app:1}

To complement and validate LBP simulations in alternating systems, we use here the Lindblad
quantum master equation (QME) and study charge transport in linear  chains.
The Lindblad equation has the form of \cite{Breuer}
\bea
\Dot{\rho} &=& -i\left[\hat H_W,\rho\right]
+ \sum_{n\geq1}
\gamma_{n}\left[\hat L_{n}\rho \hat L_{n}^\dagger-\frac{1}{2}\{\hat L_{n}\hat  L_{n}^\dagger,\rho\}\right].
\nonumber\\
\label{eq:lindblad}
\eea
Here, $\rho$ is the density matrix of the molecular electronic 
system with the Hamiltonian $\hat H_W$. 
Limiting our study to local decoherence effects, the Lindblad jump operators are selected as 
$\hat L_n=\gamma_n\ket{n}\bra{n}$,
where $n$ denotes a specified site and $\gamma_n$ is the dephasing rate constant at each site.
We organize the elements of the density matrix as a vector with the matrix $M$
prepared based on Eq. (\ref{eq:lindblad}),
$\dot{\rho}=M\rho$. 

To solve the problem in steady state and calculate the particle flux (which roughly 
corresponds to the electrical conductance in the LBP method) we 
fix the population of the entry site `1' at $\rho_{1,1}^{ss}=1$. To induce the steady state,
population is  depleted from the last site with a rate constant $\Gamma_{leak}$, 
which is an additional parameter in the problem (roughly corresponding 
to the hybridization energy in the LBP method).
In the long time limit, $\dot \rho=0$ and
the time-dependent QME becomes a linear problem with  \cite{Naim}
\bea
&\left(
\begin{matrix}
1\\
0\\
\vdots\\
0\\
\end{matrix}
\right)=
\left(
\begin{matrix}
1 & 0 & \dots & 0\\
 & & & \\
 & &  M_{m,n}& \\
  & & & \\
 &  \dots &  & M_{N,N}-\Gamma_{leak}\\
\end{matrix}
\right)
\left(
\begin{matrix}
\rho_{1,1}^{ss}\\
\rho_{1,2}^{ss}\\
\vdots\\
\rho_{N,N}^{ss}\\
\end{matrix}
\right).
\label{eq:M}
\eea
Note that $-\Gamma_{leak}\rho_{N,N}$ is added to the equation of motion of the last site; half the decay rate process 
 $(-\Gamma_{leak}/2)\rho_{i,N}$  and $(-\Gamma_{leak}/2)\rho_{N,i}$, $\forall i$,  are added to the off-diagonal terms involving the leak site \cite{NitzanB}.
In Eq. (\ref{eq:M}),  $M_{m,n}$  is constructed based on the dissipator and $\rho^{ss}$ 
are the steady state elements of the system's density matrix,  obtained after a matrix inversion.
As a proxy to the electrical conductance, the carrier's flux is given by
 $\Phi \equiv \Gamma_{leak}\rho_{N,N}^{ss}$.

\begin{figure}
\includegraphics[width = 0.5\textwidth]{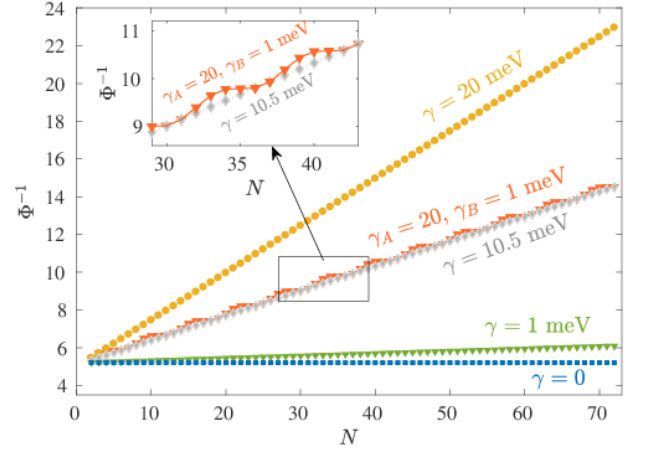} 
\caption{Inverse particle flux calculated with
the Lindblad QME at different $\gamma$, 
demonstrating that in periods of 6 sites (inset)
the flux for a system with an alternating probe condition ($\gamma_A\neq\gamma_B$)
is equal to the flux with an averaged probe value at $(\gamma_A+\gamma_B)/2$.
Simulation parameters are $\Gamma_{leak}=0.3$ and $\tau=0.2$ in eV.
When multiplying $\Phi^{-1}$ by $\hbar/|e|$ (in MKS, with $|e|$ the charge of electrons) we 
recover results in units of sec.
}
\label{fig:Lind}
\end{figure}

We simulate two scenarios: 
homogeneous chains with probe coupling $\gamma$, and a modular  wire with $\gamma_A$
and $\gamma_B$ alternating every three sites.
Results are presented in Fig. \ref{fig:Lind}. 
We find that at each period (6 sites) 
the flux of carriers in the alternating-block system is the same as
the flux calculated for the averaged-uniform system of $\gamma=(\gamma_A+\gamma_B)/2$.
Once again, this suggests that an averaging effect is taking place in the alternating-$\gamma$ 
system, similarly to what we observed with the LBP method, Sec. \ref{sec:longA}.

We note that in the model considered here all energy levels are equal,
unlike the molecular junction model with $\epsilon_n-\epsilon_F>0$.
This implies that in the present case the flux does not include contact effects.






\begin{thebibliography}{4}

\bibitem{Barton} 
A. R. Arnold, M. A.  Grodick, and J. Barton, 
DNA Charge Transport: From Chemical Principles to the Cell,
Cell Chemical Biology {\bf 23}, 183 (2016).

\bibitem{Guo2021} 
C. Yang, Z. Liu, Y. Li, S. Zhou, C. Lu, Y. Guo, M. Ramirez, Q. Zhang, 
Y. Li, Z. Liu, K. N. Houk, D. Zhang, and X. Guo, 
Electric field-catalyzed single-molecule Diels-Alder reaction dynamics,
Science Advances {\bf 7}, eabf0689 (2021).

\bibitem{Cuevas2010}
J. C. Cuevas and E. Scheer, 
{\it Molecular electronics: an introduction to theory and experiment} (World Scientific, Singapore, 2010).

\bibitem{Latha}
L. Venkataraman, J. E. Klare, C. Nuckolls, M. S. Hybertsen, and M. L.  Steigerwald,
Dependence of single-molecule junction conductance on molecular conformation,
Nature {\bf 442}, 904 (2006).

\bibitem{Frisbie1} 
S. H. Choi, B. Kim, and C. D. Frisbie, 
Electrical Resistance of Long Conjugated Molecular Wires,
 Science {\bf 320}, 1482 (2008).

\bibitem{Frisbie2} 
S. H. Choi, C. Risko, M. C. R. Delgado, B. Kim, J.-L. Bredas, and C. D. Frisbie, 
Transition from Tunneling to Hopping Transport in Long, Conjugated Oligo-imine Wires Connected to Metals, 
J. Am. Chem. Soc. {\bf 132}, 4358 (2010).


\bibitem{Ratner}
N. Renaud, Y. A. Berlin, F. D. Lewis, M. A.  Ratner,
Between Superexchange and Hopping: An Intermediate Charge-Transfer Mechanism in Poly(A)-Poly(T) DNA Hairpins,
J. Am. Chem. Soc. {\bf 135}, 3953 (2013).

\bibitem{Mccree}
H. Yan, A. J.  Bergren, R. McCreery, M. L. Della Rocca,  P. Martin, P.  Lafarge, and J. C. Lacroix,
Activationless Charge Transport across 4.5 to 22 Nm in Molecular Electronic Junctions,
 Proc. Natl. Acad. Sci. U. S. A. {\bf 110},  5326  (2013).


\bibitem{Tao}
Y. Li, L. Xiang, J. L. Palma, Y. Asai, and N. Tao,
Thermoelectric effect and its dependence on molecular length and sequence in single DNA molecules,
Nature communications {\bf 7}, 1 (2016).

\bibitem{Lacroix}
X.  Yao, X. Sun, F. Lafolet, and J.-C. Lacroix,
Long-Range Charge Transport in Diazonium-Based Single-Molecule Junctions,
Nano Lett.  {\bf 20}, 6899 (2020).

\bibitem{Carmen} 
S. Kr\"oncke and C. Herrmann,
Toward a First-Principles Evaluation of Transport Mechanisms in Molecular Wires,
J. Chem. Theory Comput.  {\bf 16}, 6267 (2020).


\bibitem{Galperin07}
M. Galperin, M. A. Ratner, and A. Nitzan,
Molecular transport junctions: vibrational effects,
Journal of Physics: Condensed Matter {\bf 19}, 103201 (2007).

\bibitem{Thoss09}
R. H\"artle, C. Benesch, and M. Thoss,
Vibrational nonequilibrium effects in the conductance of single molecules with multiple electronic states,
Phys. Rev. Lett. {\bf 102}, 146801 (2009).

\bibitem{Bijay15}
B. K. Agarwalla, J.-H. Jiang, and D. Segal,
Full counting statistics of vibrationally-assisted electronic conduction: transport and fluctuations of the thermoelectric efficiency,
Phys. Rev. B {\bf 92}, 245418 (2015).

\bibitem{Galperin20}
G. Cohen and M. Galperin,
Green’s function methods for single molecule junctions,
J. Chem. Phys. {\bf  152}, 090901 (2020).

\bibitem{Kubar09} 
P. B. Woiczikowski, T. Kubar, R. Gutiérrez, R. A. Caetano, G. Cuniberti, and M. Elstner,
Combined density functional theory and Landauer approach for hole transfer in DNA along classical molecular dynamics trajectories, 
J. Chem. Phys. {\bf 130}, 215104 (2009). 

\bibitem{Kubar1}
T. Kubar, M. Elstner, B. Popescu, and U. Kleinekath\"ofer, 
Polaron Effects on Charge Transport through Molecular Wires: A Multiscale Approach, 
J. Chem. Theory and Comp. {\bf 13}, 286 (2017).


\bibitem{Kubar2} 
M. Wolter, M. Elstner, U. Kleinekath\"oher, and T. Kubar, 
Microsecond Simulation of Electron Transfer in DNA: Bottom-Up Parametrization of an Efficient Electron Transfer Model Based on Atomistic Details, 
J. Phys. Chem. B {\bf 121}, 529 (2017).

\bibitem{Markussen17} 
T. Markussen, M. Palsgaard, D. Stradi, T. Gunst, M. Brandbyge, and K. Stokbro,
Electron-phonon scattering from Green’s function transport combined with molecular dynamics: Applications to mobility predictions, 
Phys. Rev. B {\bf 95}, 245210 (2017).

\bibitem{Ignacio18} 
L. Mejía, N. Renaud, and I. Franco, 
Signatures of Conformational Dynamics and Electrode-Molecule Interactions in the Conductance Profile During Pulling of Single-Molecule Junctions,
J. Phys. Chem. Lett. {\bf 9}, 745 (2018). 


\bibitem{Ignacio19}
Z. Li and I. Franco, 
Molecular Electronics: Toward the Atomistic Modeling of Conductance Histograms, 
J. Phys. Chem. C {\bf 123}, 9693 (2019). 


\bibitem{Ignacio21} 
Z. Li, L. Mejía, J. Marrs, H. Jeong, J. Hihath, and I. Franco, 
Understanding the Conductance Dispersion of Single-Molecule Junctions, 
J. Phys. Chem. C {\bf 125}, 3406 (2021). 


\bibitem{Buttiker1} 
M. B\"uttiker, 
Small Normal-Metal Loop Coupled to an Electron Reservoir, 
Phys. Rev. B {\bf 32}, 1846 (1985).

\bibitem{Buttiker2} 
M. B\"uttiker, 
Role of Quantum Coherence in Series Resistors,
Phys. Rev. B {\bf 33}, 3020 (1986).

\bibitem{Kilgour1}
M. Kilgour and D. Segal, 
Charge transport in molecular junctions: From tunneling to hopping with the probe technique, 
J. Chem. Phys. {\bf 143}, 024111 (2015). 

\bibitem{Korol4} 
R. Korol, M. Kilgour, and D. Segal,
ProbeZT: Simulation of transport coefficients of molecular electronic junctions under environmental effects using Buttiker's probes,
Computer Physics Communications {\bf 224}, 396 (2018).


\bibitem{Anantram2}
Y. Li, J. M. Artes, J. Qi, I. A. Morelan, P. Feldstein, M. P.  Anantram, and J. Hihath,
Comparing charge transport in oligonucleotides: RNA: DNA hybrids and DNA duplexes,
J. Phys. Chem. Lett. {\bf 7}, 1888 (2020).
%

\bibitem{Anantram1}
J. Qi, N. Edirisinghe, M. G. Rabbani, and M. P. Anantram,
Unified model for conductance through DNA with the Landauer-Büttiker formalism,
Phys. Rev.  B {\bf 87},  085404 (2013).

\bibitem{Anantram3}
W. Livernois and M. P. Anantram,
Quantum Transport in Conductive Bacterial Nanowires,
IEEE 16th Nanotechnology Materials and Devices Conference (NMDC), 1-5 (2021).
%




\bibitem{Korol1} 
R. Korol, M. Kilgour, and D. Segal, 
Thermopower of molecular junctions: Tunneling to hopping crossover in DNA, 
J. Chem. Phys. {\bf 145}, 224702 (2016).

\bibitem{Korol2} 
R. Korol and D. Segal, 
From exhaustive simulations to key principles in DNA nanoelectronics,
J. Phys. Chem. C {\bf 122}, 4206 (2018). 

\bibitem{Korol3} 
R. Korol and D. Segal,
Machine Learning Prediction of DNA Charge Transport, 
J. Phys. Chem. B {\bf 123}, 2801 (2019).

\bibitem{Kim1} 
H. Kim, M. Kilgour, and D. Segal,
Intermediate coherent-incoherent charge transport: DNA as a case study,
J. Phys. Chem. C {\bf 120}, 23951 (2016).

\bibitem{Kim2} 
H. Kim and D. Segal,
Controlling charge transport mechanisms in molecular junctions: Distilling thermally-induced hopping from coherent-resonant conduction,
 J. Chem. Phys. {\bf 146}, 164702 (2017).

\bibitem{Sun1}
A.-M. Guo and Q.-F. Sun,
Spin-Selective Transport of Electrons in DNA Double Helix,
Phys. Rev. Lett. {\bf 108}, 219102 (2012).
%

\bibitem{Sun2}
A.-M. Guo and Q.-F. Sun,
Spin-dependent electron transport in protein-like single-helical molecules,
Proc. Natl. Acad. Sci. U.S.A. {\bf 111}, 11658 (2014).

\bibitem{Sun3}
P.-J. Hu, S.-X.  Wang, X.-F. Chen, X.-H. Gao, T.-F. Fang,
A.-M. Guo, and Q.-F. Sun,
Charge Transport in a Multiterminal DNA Tetrahedron: Interplay among
Contact Position, Disorder, and Base-Pair Mismatch,
Phys. Rev. App. {\bf 17}, 024074 (2022).


\bibitem{Kilgour2} 
M. Kilgour and D. Segal,
Inelastic effects in molecular transport junctions: The probe technique at high bias,
J. Chem. Phys. {\bf 144}, 124107 (2016).

\bibitem{Kilgour3}
M. Kilgour and D. Segal, Tunneling diodes with environmental effects,
J. Phys. Chem. C {\bf 119}, 25291 (2015).


\bibitem{Malay} 
M. Bandyopadhyay and D. Segal, 
Quantum heat transfer in harmonic chains with self consistent reservoirs: Exact numerical simulations, 
Phys. Rev. E {\bf 84}, 011151 (2011).


\bibitem{Roya} 
R. M. Fereidani and D. Segal, 
Phononic heat transport in molecular junctions: quantum effects and vibrational mismatch, 
J. Chem. Phys. {\bf 150}, 024105 (2019).

\bibitem{Goold21} 
A. M. Lacerda, J. Goold, and G. T.  Landi,
Dephasing enhanced transport in boundary-driven, quasiperiodic chains,
Phys. Rev. B {\bf 104}, 174203 (2021).


\bibitem{BijayLBP}
M. Saha, B. P. Venkatesh, and B. K. Agarwalla,
Quantum transport in quasi-periodic lattice systems in presence of B\" uttiker probes,
arXiv:2202.14033 (2022).


\bibitem{Pastawski}
J. L. D’Amato and H. M. Pastawski, 
Conductance of a disordered linear chain including inelastic scattering events,
Phys. Rev. B {\bf 41}, 7411 (1990).

\bibitem{Dhar07}
D. Roy and A. Dhar,
Electron transport in a one dimensional conductor with inelastic scattering by self-consistent reservoirs,
Physical Review B {\bf 75}, 195110 (2007).

\bibitem{Kiminori1}
K. Hattori and M. Yoshikawa,
Generalized self-consistent reservoir model for normal and anomalous heat transport in quantum harmonic chains,
Phys. Rev. E {\bf 99}, 062104 (2019).


\bibitem{Kiminori2}
K. Hattori and M. Sambonchiku,
Dimensional crossover of thermal transport in quantum harmonic lattices coupled to self-consistent reservoirs,
Phys. Rev. E {\bf 102}, 012121 (2020).



\bibitem{bond2003} 
D. R. Bond and D. R. Lovley,
Electricity Production by Geobacter sulfurreducens Attached to Electrodes, 
Applied and Environmental Microbiology {\bf 69}, 1548 (2003).

\bibitem{lovley2019} 
D. R. Lovley and D. J. F.Walker, 
Geobacter Protein Nanowires, 
Frontiers in Microbiology {\bf 10}, 2078 (2019).


\bibitem{Butt}
J. H. van Wonderen, {\it et al.}, 
Nanosecond heme-to-heme electron transfer rates in a multiheme cytochrome nanowire reported by a spectrally unique His/Met-ligated heme,
Proc. Natl. Acad. Sci. U.S.A. {\bf 118}, e2107939118 (2021).

\bibitem{BeratanPNAS}
D. N. Beratan, 
Multiple hops move electrons from bacteria to rocks, 
Proc. Natl. Acad. Sci. U.S.A.  {\bf 118}, 42, (2021).

\bibitem{Gu20}
Y. Gu, V. Srikanth, A. I. Salazar-Morales, R. Jain, J. P. O’Brien,
S. M. Yi, R. K. Soni, F. A. Samatey, S. E. Yalcin, and 
N. S. Malvankar,
Structure of Geobacter pili reveals secretory rather than nanowire behaviour,
Nature {\bf 597}, 430 (2021).

\bibitem{Gu20Rev}
T. Boesen, L. P. Nielsen, and A. Schramm,
Pili for nanowires,
Nature Microbiology {\bf 6}, 1347 (2021).

\bibitem{agam2020} 
Y. Agam, R. Nandi, A. Kaushansky, U. Peskin, and N. Amdursky, 
The porphyrin ring rather than the metal ion 
dictates long-range electron transport across proteins coherence-assisted mechanism, 
Proc. Natl. Acad. Sci. USA {\bf 117}, 32260 (2020).

\bibitem{Beratan} 
X. Ru, P. Zhang, and D. Beratan, 
Assessing Possible Mechanisms of Micrometer-Scale Electron Transfer in Heme-Free Geobacter sulfurreducens Pili,
J. Phys. Chem. B {\bf 123}, 5035 (2019).

\bibitem{Amdursky} 
Y. Eshel, U. Peskin, and N. Amdursky, 
Coherence-assisted electron diffusion across the multi-heme protein-based bacterial nanowire, 
Nanotechnology {\bf 31}, 314002 (2020).

\bibitem{Plenio} 
A. Mattioni, F. Caycedo-Soler, S. F. Huelga, and M. B. Plenio, 
Design principles for long-range energy transfer at room temperature, 
Phys. Rev. X {\bf 11}, 041003 (2021).


\bibitem{Aspuru17}
S. K. Saikin, M. A. Shakirov, C. Kreisbeck, U. Peskin, Y. N. Proshin, and A. Aspuru-Guzik,
On the long-range exciton transport in molecular systems: The application to H-aggregated heterotriangulene chains, 
J. Phys. Chem. C {\bf 121}, 24994 (2017).

\bibitem{BeratanE}
P. P. Roy, {\it et al.,}
Synthetic Control of Exciton Dynamics in Bioinspired Cofacial Porphyrin Dimers,
J. Am. Chem. Soc. {\bf 144}, 6298 (2022).

\bibitem{Barton34}
J. D. Slinker, N. B. Muren, S. E. Renfrew, and  J. K. Barton,
DNA charge transport over 34 nm,
Nature Chemistry {\bf 3}, 228 (2011).

\bibitem{Danny1}
G. I. Livshits, {\it et al.,}
Long-range charge transport in single G-quadruplex DNA molecules,
Nature Nanotechnol. {\bf 9}, 1040 (2014).


\bibitem{Danny2} 
R. Zhuravel,  {\it et al.,}
Backbone charge transport in double-stranded DNA,
Nature Nanotechnol. {\bf 15}, 836 (2020).


\bibitem{NitzanB} 
A. Nitzan,
{\it Chemical Dynamics in Condensed Phases: Relaxation, Transfer and Reactions in Condensed Molecular Systems},
(Oxford University Press, Oxford, UK, 2006).


\bibitem{BeratanF}
Y. Zhang, C. Liu, A. Balaeff, S. S. Skourtis, and D. N. Beratan,
Biological charge transfer via flickering resonance,
Proc. Natl. Acad. Sci. U.S.A.  {\bf 111}, 10049 (2014).

\bibitem{PeskinU}
A. D. Levine, M. Iv, and U. Peskin,
Formulation of long-range transport rates through molecular bridges: from unfurling to hopping,
J. Phys. Chem. Lett. {\bf 9}, 4139 (2018).

\bibitem{VRH}
N. Mott, 
{\it Conduction in Non-Crystalline Materials}, 
(Clarendon Press:
Oxford, U.K., 1987). 

\bibitem{Polaron}
H. Grabert and U. Weiss,
Quantum tunnelling rates for asymmetric double-well systems with Ohmic dissipation. 
Phys. Rev. Lett. {\bf 54}, 1605 (1985).

\bibitem{Taomix1} 
L. Xiang, J. L. Palma, C. Bruot, V. Mujica, M. A. Ratner, and N. Tao,
Intermediate tunnelling–hopping regime in DNA charge transport,
Nature chemistry {\bf 7}, 221 (2015).

\bibitem{Taomix2}
    C. Liu, L. Xiang, Y. Zhang, P. Zhang, D. N. Beratan, Y. Li and N. Tao,
Engineering nanometre-scale coherence in soft matter,
Nature Chemistry {\bf 8}, 941 (2016).


\bibitem{Hanggi}
P. H\"anggi, P. Talkner, and M. Borkovec,
Reaction-rate theory: fifty years after Kramers,
Rev. Mod. Phys. {\bf 62}, 251 (1990).

\bibitem{Breuer} 
H. P. Breuer  and F. Petruccione,
{\it The Theory of Open Quantum Systems}
(Oxford University Press, Oxford, UK, 2007).

\bibitem{Naim}
N. Kalantar and D. Segal,
Mean First-Passage Time and Steady-State Transfer Rate in Classical Chains,
J. Phys. Chem. C  {\bf 123}, 2, 1021 (2019).


\end{thebibliography}
\end{document}